\documentclass[reprint,amsmath,amssymb,aps,superscriptaddress,floatfix]{revtex4-2}

\usepackage[utf8]{inputenc}
\usepackage{graphicx}
\usepackage{dcolumn}
\usepackage{bm}
\usepackage{amsmath,amssymb}
\usepackage{amsthm}
\usepackage{mathrsfs}
\usepackage{indentfirst}
\usepackage{float}
\usepackage{braket}
\usepackage{physics}
\usepackage{mathtools}
\usepackage{color,graphicx} 
\usepackage{bbold}
\usepackage{subfiles}
\usepackage{subfigure}
\usepackage{hyperref}
\hypersetup{
 colorlinks=true,
 citecolor=red,
 linkcolor=blue,
 urlcolor=blue,
 pdfpagemode=UseNone,
 pdfstartview=FitH}

\newcommand{\rr}[1]{\mathrm{#1}}
\newcommand{\cl}[1]{\mathcal{#1}}

\begin{document}

\preprint{APS/123-QED}
\title{Universal Quantum Computation in Globally Driven Rydberg Atom Arrays}
\author{Francesco~Cesa}
\email{francesco.cesa@phd.units.it}
\affiliation{
Institute for Theoretical Physics, University of Innsbruck, Innsbruck A-6020, Austria}
\affiliation{Institute for Quantum Optics and Quantum Information, Austrian Academy of Sciences, Innsbruck A-6020, Austria}
\affiliation{
Department of Physics, University of Trieste, Strada Costiera 11, 34151 Trieste, Italy}
\affiliation{Istituto Nazionale di Fisica Nucleare, Trieste Section, Via Valerio 2, 34127 Trieste, Italy}

\author{Hannes~Pichler}
\email{hannes.pichler@uibk.ac.at}
\affiliation{
Institute for Theoretical Physics, University of Innsbruck, Innsbruck A-6020, Austria}
\affiliation{Institute for Quantum Optics and Quantum Information, Austrian Academy of Sciences, Innsbruck A-6020, Austria}

\date{\today}
\begin{abstract}
 We develop a model for quantum computation with Rydberg atom arrays, which only relies on global driving, without the need of local addressing of the qubits: any circuit is executed by a sequence of global, resonant laser pulses on a static atomic arrangement. We present two constructions: for the first, the circuit is imprinted in the trap positions of the atoms and executed by the pulses; for the second, the atom arrangement is circuit-independent, and the algorithm is entirely encoded in the global driving sequence. Our results show in particular that a quadratic overhead in atom number is sufficient to eliminate the need for local control to realize a universal quantum processor. We give explicit protocols for all steps of an arbitrary quantum computation, and discuss strategies for error suppression specific to our model. Our scheme is based on dual-species processors with atoms subjected to Rydberg blockade constraints, but it might be transposed to other setups as well.
\end{abstract}
\maketitle

\begin{figure}[t!]
   \centering
   \includegraphics[width=1.0\linewidth]{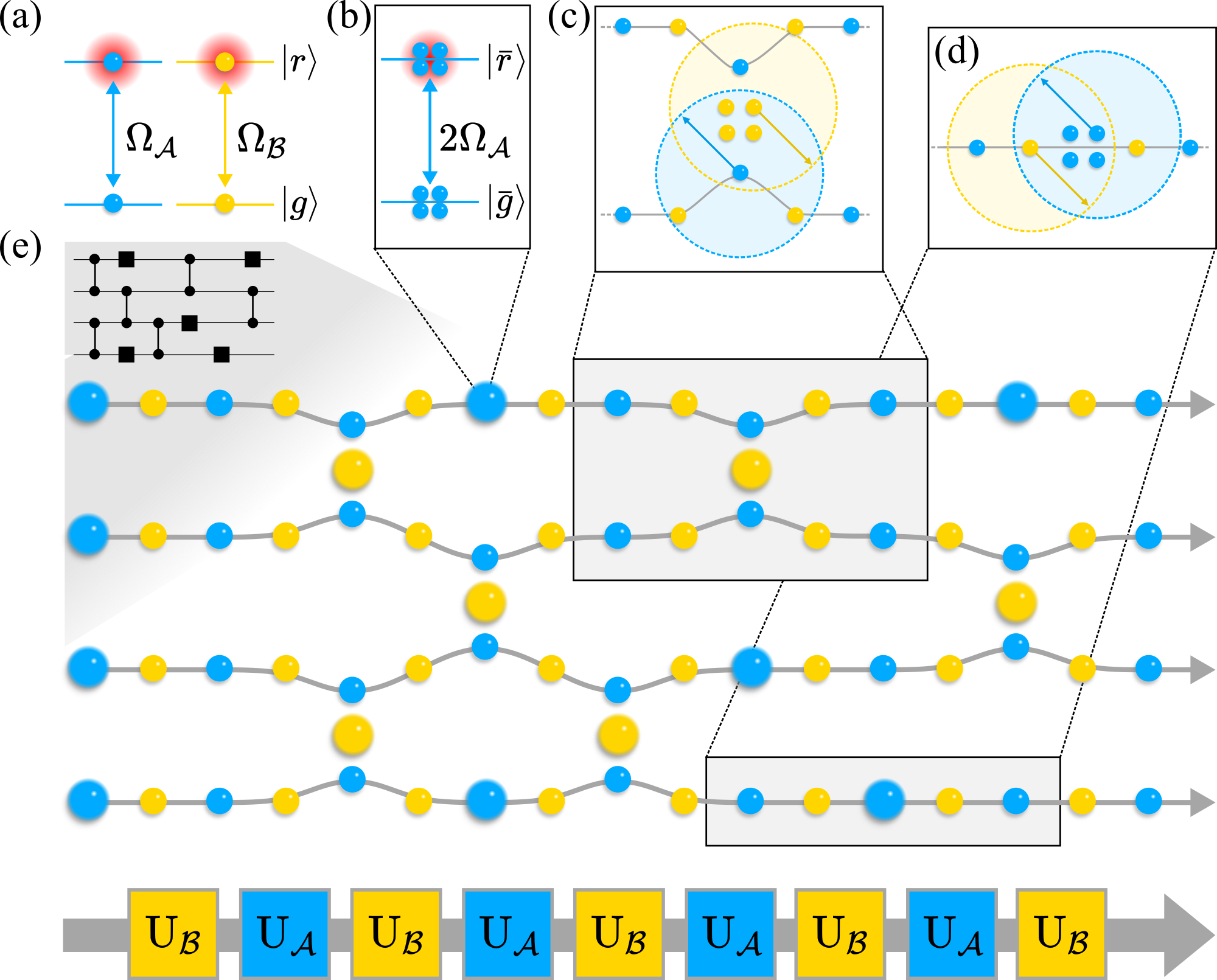}
   \caption{(a)~Two atomic species, $\cl{A}$ and $\cl{B}$. (b)~Superatoms are used as impurities. (c, d)~Blockade relations near to impurities, enabling two-~and single-qubit gates. (e)~A circuit is translated in an atomic arrangement; optical pulses propagate the information flow through the wires, which represent logical qubits. }
    \label{fig_1}
\end{figure}

\begin{figure}[t!]
   \centering
    \includegraphics[width=0.95\linewidth]{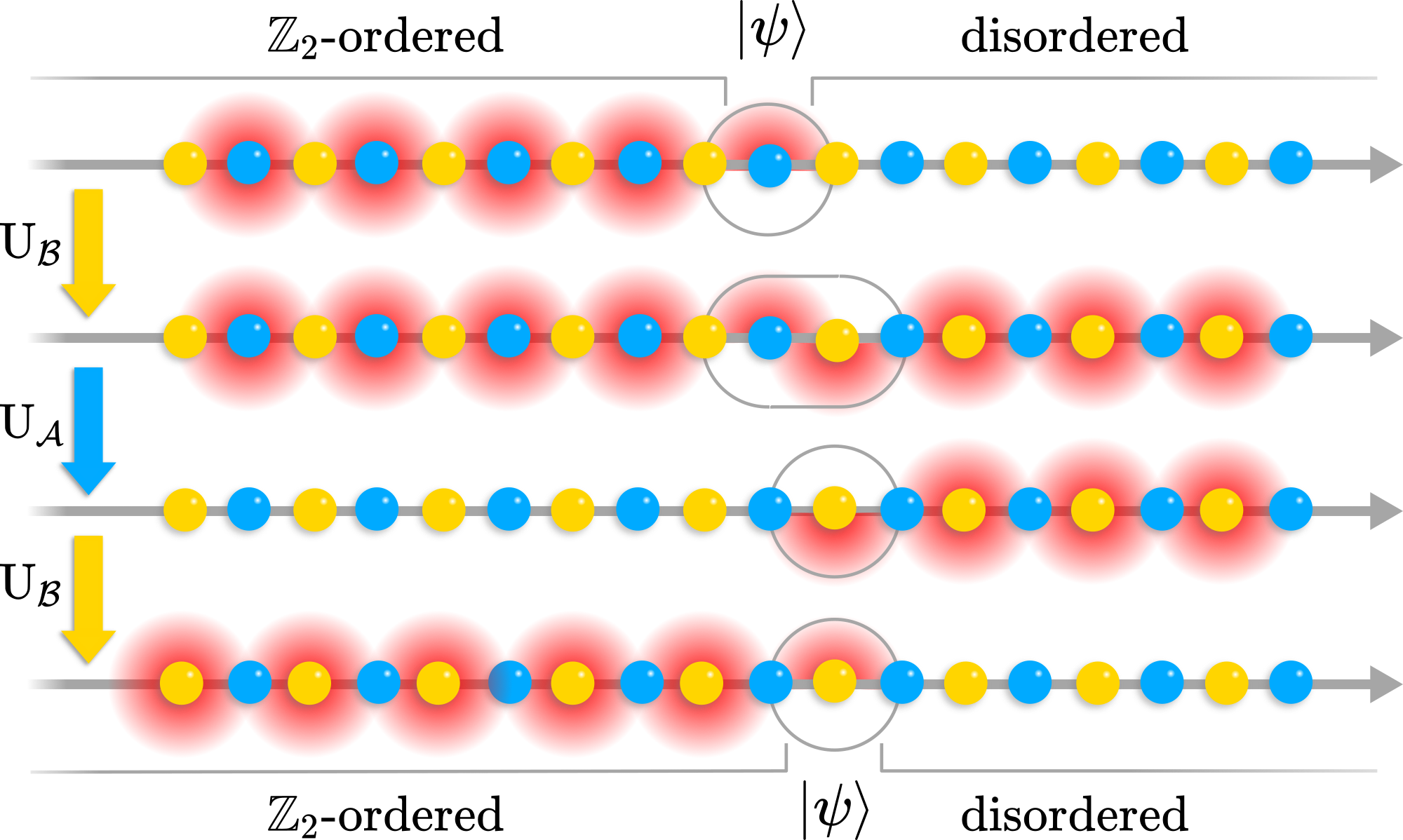}
    \caption{The state $\ket{\psi}$ is transported through a wire via global pulses. The first line identifies the \textit{standard configuration} $\ket{\Psi(k)}$, with $\ket{\psi}$ at the interface between a $\mathbb{Z}_2-$ordered and a disordered sector. The $\rr{U}_\cl{X}$ apply blockade-conditioned flips of the atomic state.}
    \label{fig_2}
\end{figure}

\indent \textit{Introduction.---} Quantum computation (QC) leverages quantum effects for solving computational problems. In the standard paradigm~\cite{NIELSEN_CHUANG}, known as the circuit model (CM), an  algorithm is executed by applying a sequence of gates on a register of quantum units (qubits); alternatively, other models have been developed, including measurement-based~\cite{ONE_WAY}, adiabatic~\cite{FAHRI_ADIABATIC, AHARONOV_ADIABATIC_PROOF,LIDAR_ADIABATIC}, and topological~\cite{SARMA_TOPOLOGICAL} schemes. While they are all fundamentally equivalent, particular features of each model are appealing for specific implementations.\\ 
\indent Recently, neutral atoms in optical tweezers have emerged as a promising platform for quantum information processing~\cite{SAFFMAN_MOLMER, browaeysManybodyPhysicsIndividually2020, kaufmanQuantumScienceOptical2021}: deterministically assembled arrays are coherently manipulated with optical pulses and entangled through Rydberg interactions~\cite{jakschFastQuantumGates2000, lukinDipoleBlockadeQuantum2001, wilkEntanglementTwoIndividual2010_2, levineParallelImplementationHighFidelity2019, madjarovHighfidelityEntanglementDetection2020a, ebadiQuantumOptimizationMaximum2022a, steinert2022spatially}, to realize programmable and scalable processors~\cite{BLUVSTEIN_PROCESSOR, maUniversalGateOperations2022, grahamMultiqubitEntanglementAlgorithms2022}. One approach for QC is then to realize the CM: qubits are encoded in atomic levels, with gates executed via \textit{local} optical control. While this approach has seen remarkable progress~\cite{grahamMultiqubitEntanglementAlgorithms2022, scholl2023erasure, evered2023highfidelity, ma2023highfidelity}, individual addressing of atoms remains a formidable technological challenge.\\ 
\indent Here, we develop an alternative model for QC with Rydberg atoms, which does not require local addressing or dynamical rearrangement. Instead, the quantum algorithm is executed by driving a static atomic array with a \emph{global}, resonant laser in the Rydberg blockade regime. Specifically, a universal quantum processor featuring $n$ logical qubits is realized with an arrangement of $\cl{O}(n^2)$ atoms; any circuit of depth $p$ is then executed by a sequence of $\cl{O}(np)$ pulses driving the transition between an internal electronic ground and a Rydberg state. All atoms are initialized in the internal ground state, and the final read-out is in this computational basis, requiring no local control. We present our model by first translating any quantum algorithm to a sequence of global pulses on a circuit-dependent arrangement of $\cl{O}(n^2p)$ atoms (see Fig.~\ref{fig_1}e); then, we reduce this to a \emph{universal} arrangement of $\cl{O}(n^2)$ atoms. We demonstrate this novel QC model on the example of a dual-species array with species-selective global drivings~\cite{BERNIEN_MID_CIRCUIT_SCIENCE, PRL_ISOTOPES_1, SEMEGHINI_BULLETIN_1, SEMEGHINI_BULLETIN_2, KANG-KUEN_NI, singhDualElementTwoDimensionalAtom2022, PRL_ISOTOPES_2}, but it can be transposed to other architectures with static hardware layout.\\
\indent \textit{Main picture.---} Our QC model is based on several key ideas, which we outline in the following, before detailing them individually. We consider $n$ 1D wires of atoms, each hosting a qubit degree of freedom. At each step, the qubit state is located at a single atom $k$, particularly, at the interface between an ordered and a disordered sector of the wire (Fig.~\ref{fig_2}, top); we denote this as $\ket{\Psi(k)}$. We design a sequence of global laser pulses, such that
\begin{equation}\label{traslation}
    \ket{\Psi(k)}\xrightarrow[]{\text{pulses}}\ket{\Psi(k+1)};
\end{equation}
essentially, the qubit propagates through the wire as the pulses progress - defining an information flow. \\
\indent By breaking the symmetry at selected points, we manipulate the quantum information during this propagation; specifically, we introduce two types of impurities. We insert impurities of the first type \emph{inside} the wires: when an interface reaches them, we apply sequences of global pulses, which effectively manipulate individually the impurities at the interfaces (single-qubit gates). We insert impurities of the second type \emph{between} the wires, triggering interactions between the interfaces when they pass by (multiple-qubit gates). The combination of these constructions translates any algorithm into an arrangement of atoms, where wires represent the lines in the circuit, and impurities the location of gates (Fig.~\ref{fig_1}e). In the remainder we elaborate on these ideas, as well as on the remaining aspects required for universal QC, such as initialization.\\
\indent \textit{Physical setup.---}  We consider a tweezer array with two atomic species, $\cl{A}$ and $\cl{B}$, where atoms can be arranged in arbitrary 2D configurations~\cite{kaufmanQuantumScienceOptical2021, labuhnTunableTwodimensionalArrays2016,ENDRES_ATOM_BY_ATOM, BARREDO_SCIENCE, semeghiniProbingTopologicalSpin2021_2, PRL_ISOTOPES_2, singhDualElementTwoDimensionalAtom2022}. For each atom, we consider two electronic states (Fig.~\ref{fig_1}a): a ground state $\ket{g}$, and a highly excited Rydberg state $\ket{r}$; we define the Pauli algebra as $\rr{Z}=\ket{g}\!\bra{g}-\ket{r}\!\bra{r}$, $\rr{X}=\ket{r}\!\bra{g}+\ket{g}\!\bra{r}$. The two levels can be coherently coupled in a species-selective way with global resonant fields, i.e. lasers acting identically on all atoms of the same species. Moreover, atoms in $\ket{r}$ interact pairwise via induced dipole-dipole interactions. The Hamiltonian is $\cl{H}=\cl{H}_\cl{A}+\cl{H}_\cl{B}+\cl{H}_\text{int}$, with
\begin{equation}\label{hamiltonian}
\begin{split}
& \cl{H}_{\cl{X}} = \frac{\hbar\Omega_{\cl{X}}}{2}\sum_{i\in\cl{X}}\left[e^{i\phi_\cl{X}}\ket{g_i}\!\bra{r_i}+ \text{h.c.}\right], \\
    & \cl{H}_\text{int}=\frac{1}{2}\sum_{i\neq j}V_{i,j}\!\ket{r_ir_j}\!\bra{r_ir_j}.
\end{split}
\end{equation}
Here, $\Omega_\cl{X}$ and $\phi_\cl{X}$ are the Rabi frequency and phase of the laser driving $\cl{X}\in\left\{\cl{A},\cl{B}\right\}$, and $V_{i,j}$ is the interaction strength  between atoms $i$ and $j$. This includes both intra- and inter-species interactions~\cite{SAFFMAN_FORSTER_PRA}. The interaction strength decays with the atomic separation, $d_{i,j}$, as $V_{i,j}\!\sim\!d_{i,j}^{-6}$. At short distances this gives the Rydberg blockade effect~\cite{urbanObservationrydberg,gaetanObservationCollectiveExcitation2009, BARREDO_PRL}, preventing more than one excitation within a blockade radius $R_B$, while at larger separations interactions are negligible. We assume for every pair of atoms either perfect blockade or no interactions~\cite{supplemental}, resulting in a dual-species PXP-model~\cite{serbyn2021quantum, PhysRevB.66.075128}. The parameters $\Omega_{\cl{X}}$ and $\phi_{\cl{X}}$ are controlled time-dependently, but the atomic arrangements, and thus the mutual blockade relations, are static.

\begin{figure*}[t!]
    \begin{minipage}{1.0\textwidth}
        \includegraphics[width=\textwidth]{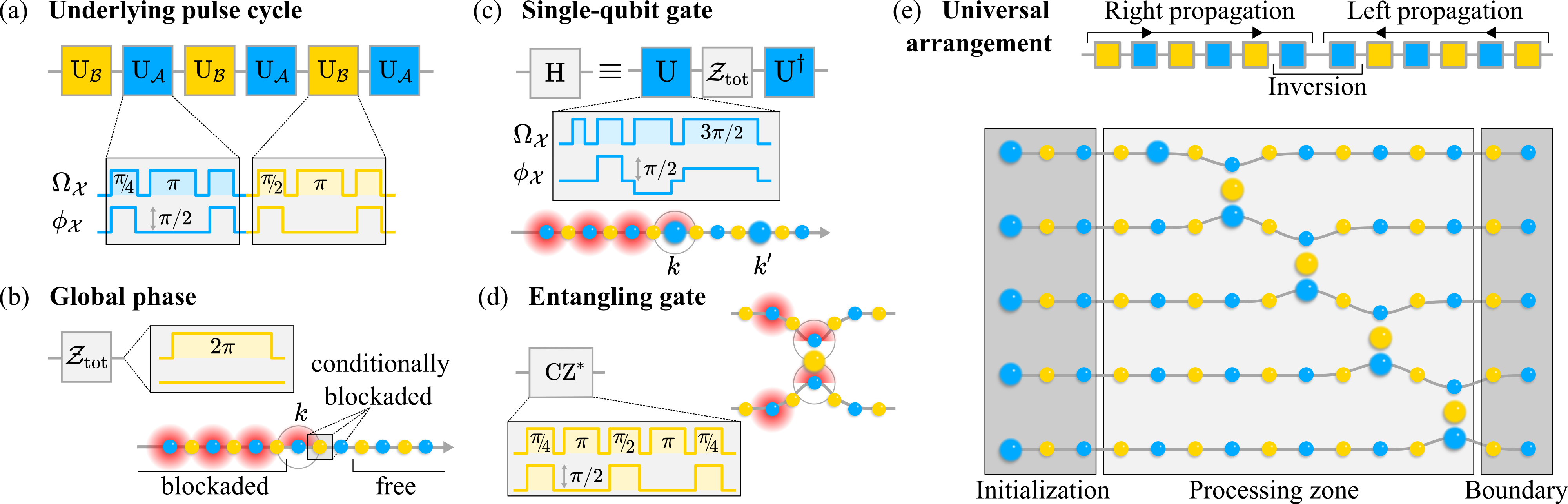}
    \end{minipage}%
\caption{(a) The pulses $\rr{U}_\cl{X}$ act as (blockade-conditioned) flips of the atomic state. The first line shows the duration of the pulses (units of $1/\Omega_\cl{X}$), while the second the phase of the laser. (b-d) By globally driving the whole array, exploiting impurities we achieve single- and two-qubit gates ($\rr{CZ}^*=\cl{Z}_\text{tot}\rr{CZ}$). (e) Universal arrangement: moving the information both to the left and to the right, at each step we localize it at positions where we achieve the desired individual control.}
\label{fig_3}
\end{figure*}

\textit{Logical qubits.---} We start by describing our basic building block: a single atomic wire hosting a logical qubit. The two species alternate, and are subjected to nearest-neighbor blockade; we label the atoms by $k\in\left\{0,1,2,...\right\}$.\\
\indent A wire is in the \emph{standard configuration} (SC), with the \emph{interface} at site $k$, if all atoms at $k'<k$ are in the $\mathbb{Z}_2-$ordered configuration, $\ket{\dots r_{k\!-\!4}g_{k\!-\!3}r_{k\!-\!2}g_{k\!-\!1}}$, while for $k'>k$ they are all in the ground state, $\ket{g_{k\!+\!1}g_{k\!+\!2}g_{k\!+\!3}\dots}$; the interface atom at $k$ can be in any superposition $\ket{\psi_k}=\alpha\ket{g_k}+\beta\ket{r_k}$. We use the notation
\begin{equation}\label{standard}
    \ket{\Psi(k)}=\ket{...r_{k\!-\!4}g_{k\!-\!3}r_{k\!-\!2}g_{k\!-\!1}}\ket{\psi_k}\ket{g_{k\!+\!1}g_{k\!+\!2}g_{k\!+\!3}...},
\end{equation}
depicted in Fig.~\ref{fig_2}, top.\\
\indent Throughout the computation, we apply a sequence of pulses of the form
\begin{equation}\label{clock}
   ...\rr{U}_\cl{B}\rr{U}_\cl{A}\rr{U}_\cl{B}\rr{U}_\cl{A}\rr{U}_\cl{B}\rr{U}_\cl{A}...,
\end{equation}
alternating the addressed species. Here $\rr{U}_\cl{X}$ induce a conditional state-flip of the $\cl{X}-$atoms, flipping the populations, $\ket{g}
\leftrightarrow\ket{r}$, if all neighbors are in $\ket{g}$. While this might be implemented by $\pi$ pulses, we employ a generalized pulse design (Fig.~\ref{fig_3}a), for reasons that will become clear below. \\
\indent Without loss of generality assume the atom hosting the interface at site $k$ is of the $\cl{A}-$type; we now show that~\footnote{\indent This holds away from the boundaries, and specifically~\eqref{standard} is intended for $k\geq 2$. }
\begin{equation}\label{transport}
 \rr{U}_\cl{B} \rr{U}_\cl{A} \rr{U}_\cl{B}\ket{\Psi(k)} = \ket{\Psi(k+1)}.
\end{equation}
This can be understood by considering the blockade constraints step-by-step (Fig.~\ref{fig_2}). First we apply $\rr{U}_\cl{B}$: because the system is in the SC~\eqref{standard}, all $\cl{B}-$atoms to the left of $k$ are blockaded, while those to the right get excited; only the atom at site $k+1$ evolves non-trivially, conditioned on the state of the qubit at $k$. It gets excited iff the latter is in $\ket{g_k}$, leading to 
\begin{equation}
     \ket{...g_{k\!-\!3}r_{k\!-\!2}g_{k\!-\!1}}\!\Big[ \alpha\!\ket{gr}\!+\! \beta\!\ket{rg} \Big]_{k,k\!+\!1}\!\ket{g_{k\!+\!2}r_{k\!+\!3}g_{k\!+\!4}...},
\end{equation}
with $\ket{\psi}$ shared between $k$ and $k+1$ (Fig.~\ref{fig_2}, second line). Then, $\rr{U}_\cl{A}$ de-excites the $\cl{A}-$atoms at $k'\leq k$, while all others are blockaded, yielding
\begin{equation}
     \rr{U}_\cl{A}\rr{U}_\cl{B}\!\ket{\Psi(k)}\!=\!\ket{...g_{k\!-\!2}g_{k\!-\!1}\!g_k}\!\Big[ \rr{X}\!\ket{\psi} \Big]_{k\!+\!1}\!\ket{g_{k\!+\!2}r_{k\!+\!3}g_{k\!+\!4}...},
\end{equation}
moving the (flipped) qubit to $k+1$  (Fig.~\ref{fig_2}, third line). Finally, $\rr{U}_\cl{B}$ excites the $\cl{B}-$atoms at $k'\leq k$ and de-excites those at $k'\geq k+2$, while flipping $k+1$, leading to Eq.~\eqref{transport} (Fig.~\ref{fig_2}, bottom line).\\
\indent Altogether, the cycle~\eqref{clock} moves the SC~\eqref{standard} as in Eq.~\eqref{transport}: it induces a dispersionless flow of quantum information, with the interface moving to the right. This is the underlying mechanism of our model: while the pulse cycle propagates the information, we manipulate it by inserting additional global pulses at stroboscopic times; impurities will enable non-trivial operations. The rest of the paper is devoted to explaining how we can unlock the full toolbox of universal QC. From now on we consider $n$ wires (Fig.~\ref{fig_1}e), each hosting its qubit; the pulse cycle thus moves all qubits in parallel through their wires.\\  
\indent \textit{Collective phase gates.---} One key operation imprints a phase on all the qubits in parallel, specifically, $\cl{Z}_\text{tot}=\bigotimes_\cl{Q}\rr{Z}_\cl{Q}$, where $\cl{Q}$ runs over \emph{logical} qubits (wires). Consider the SC with the interface on an $\cl{A}-$atom at $k$ (Fig.~\ref{fig_3}b). To realize $\cl{Z}_\text{tot}$, we drive $\cl{B}$ with a global $2\pi$ pulse: atoms to the left of $k$ are blockaded, and those at $k'>k+1$ acquire an irrelevant global phase. However, the $\cl{B}-$atom at site $k+1$ picks up a phase $-1$ iff the qubit at $k$ is in $\ket{g_k}$. This conditional phase is thus equivalent to the application of a $\rr{Z}$ gate at $k$; this happens in parallel for all the wires, yielding $\cl{Z}_\text{tot}$. \\  
\indent\textit{Impurities.---} A \emph{superatom} is a cluster of $S$ atoms, all arranged within a radius~$ R_B$. The blockade constrains them to a collective dynamics, where only one can be excited at a time; this effectively realizes a two-level system, featuring a ground state $\ket{\bar{g}}$, with all cluster-atoms in $\ket{g}$, and an effective excited state $\ket{\bar{r}}$, with a single excitation shared among the cluster. Under the Hamiltonian~\eqref{hamiltonian} the Rabi frequency of this two-level system is enhanced as $\sqrt{S}\Omega_\cl{X}$. While the arguments below transpose to any $S\geq 2$, we consider $S=4$, with the prescription
\begin{equation}
    \ket{\bar{g}}\!\equiv\!\ket{gggg},\;\;\ket{\bar{r}}\!\equiv\!\frac{\ket{rggg}\!+\!\ket{grgg}\!+\!\ket{ggrg}\!+\!\ket{gggr}}{2},
\end{equation}
and the corresponding Pauli algebra. We assume each superatom is always either completely inside or outside the blockade radii of other atoms, and thus behaves as a unique entity.\\
\indent Since the superatoms have different Rabi frequency, the pulses realizing $\rr{U}_\cl{X}$ must be carefully designed, not to interfere with the previous protocols (Fig.~\ref{fig_3}a). Below we replace some $\cl{A}-$atoms inside the wires with $\cl{A}-$superatoms. Thus, $\rr{U}_\cl{A}$ is designed to realize a flip of the atomic state for both. Differently, $\cl{B}-$superatoms will be  introduced between the wires, and we require $\rr{U}_\cl{B}$ to act as a $4\pi$ rotation on on them; also the $2\pi$ pulse for $\cl{Z}_\text{tot}$ translates to a $4\pi$.

\indent \textit{Single-qubit unitaries.---} If we place ans $\cl{A}-$superatom instead of an $\cl{A}-$atom at site $k$ along a wire, we can execute a gate on the corresponding qubit when the interface is located at $k$. For universal QC we need unitaries of the form $\rr{U}(\phi,\alpha)=e^{-i\alpha\rr{R}(\phi)/2}$, where $\rr{R}(\phi)=\cos(\phi)\rr{X}+\sin(\phi)\rr{Y}$ (rotations around axes in the $xy$ plane); rotations around $\hat{z}$ are accounted for by classical processing~\footnote{Since $\rr{U(\phi,\alpha)}e^{-i\beta\rr{Z}/2}=e^{-i\beta\rr{Z}/2}\rr{U(\phi-\beta,\alpha)}$, if any unitary of the form $\rr{U(\phi,\alpha)}$ can be physically implemented, phase gates can always be propagated to the end of the circuit, where they can be ignored as measurements are performed in the computational basis.}. Thus, we will ignore phase corrections, and denote by $\stackrel{z}{=}$ equalities holding up to phases.\\
\indent When the interface is located at the superatom, we apply a global pulse sequence, designed for the desired unitary, such that:~(i) normal atoms
(of both species), and $\cl{B}-$superatoms, wherever they are
located, are unaffected;~(ii) an $\cl{A}-$superatom placed at $k$ undergoes the target unitary;~(iii) $\cl{A}-$superatoms placed at $k'\neq k$
are unaffected. We first show how to engineer a Hadamard, $\rr{H}\!=\!\left(\rr{X}\!+\!\rr{Z}\right)\!/\!\sqrt{2}\stackrel{z}{=}e^{-i\pi\rr{Y}/4}$, and then generalize to arbitrary gates.\\
\indent The pulse sequence for the Hadamard is shown in Fig.~\ref{fig_3}c. It consists of three steps: first, a pulse sequence on $\cl{A}$, inducing a single-atom evolution $\rr{U}$; then, the $\cl{Z}_\text{tot}$ protocol; finally, the inverse of the first, $\rr{U}^\dag$. The first step acts on $\cl{A}-$atoms as~\footnote{$\rr{U}=-e^{i\pi\rr{Z}/8}$.}
\begin{equation}\label{normal_atoms}
    \rr{U} = \rr{U}\!\left( \!\frac{\pi}{4},\frac{3\pi}{2}\! \right) \rr{U}\!\left(\! -\frac{\pi}{4},\frac{3\pi}{4} \! \right) \rr{U}\!\left(\! \frac{\pi}{2},\frac{\pi}{2}\! \right) \rr{U}\!\left(\! 0,\frac{\pi}{4}\! \right)\stackrel{z}{=}\mathbb{1};
\end{equation}
since it does not affect the atomic populations, during the second step all blockade constraints are maintained, so that $\cl{Z}_\text{tot}$ works as described above. Hence, for $\cl{A}-$atoms the sequence acts as $\rr{U}^\dag\rr{Z}\rr{U}=\rr{Z}\stackrel{z}{=}\mathbb{1}$ (for atoms at the interface), or $\rr{U}^\dag\rr{U}=\mathbb{1}$ (otherwise), thus addressing~(i).\\
\indent Consider now the superatom at $k$. Due to the enhanced Rabi frequency, the first (third) step induces a different unitary $\bar{\rr{U}}$ ($\bar{\rr{U}}^\dag$). Specifically, the evolution is found by considering effective doubled pulse durations in place of Eq.~\eqref{normal_atoms}, obtaining~\footnote{$\bar{\rr{U}}=-ie^{i\pi\rr{Z}/8}\rr{X}e^{i\pi\rr{Y}/8}$.}
\begin{equation}\label{superatoms}
    \bar{\rr{U}} = \rr{U}\!\left( \!\frac{\pi}{4},3\pi\! \right) \rr{U}\!\left(\! -\frac{\pi}{4},\frac{3\pi}{2} \! \right) \rr{U}\!\left(\! \frac{\pi}{2},\pi\! \right) \rr{U}\!\left(\! 0,\frac{\pi}{2}\! \right);
\end{equation}
this corresponds to a $\pi/4$ rotation. Then, because $k$ is the interface of the SC, the second step acts as a $\bar{\rr{Z}}$ gate on it (as part of $\cl{Z}_\text{tot}$); finally the inverse of the first step is applied. Eventually, one finds  
\begin{equation}
    \bar{\rr{U}}_\text{tot}=\bar{\rr{U}}^\dag \bar{\rr{Z}} \bar{\rr{U}} = -\frac{\bar{\rr{X}}+\bar{\rr{Z}}}{\sqrt{2}}\;\stackrel{z}{=}\;-e^{-i\pi\bar{\rr{Y}}/4},
\end{equation}
i.e., the overall operation is a Hadamard. Thus~(ii) is fulfilled. Regarding~(iii), any $\cl{A}-$superatom located at $k'\neq k$ does not feel $\cl{Z}_\text{tot}$ because it is not at the interface of the SC: if $k'\leq k-2$ its neighboring $\cl{B}-$atoms are both blockaded; if $k'\geq k+2$ none is, hence the \textit{two} conditional phases annihilate as $\bar{\rr{Z}}^2=\mathbb{1}$. Therefore, the first and third step cancel out, leading to an identity. \\   
\indent The above arguments generalize to $\pi/2$ rotations around any axis in the $xy$ plane: any $\rr{U}(\phi,\frac{\pi}{2})$ is achieved by shifting all the phases by $\phi-\pi/2$ in Eqs.~\eqref{normal_atoms} and~\eqref{superatoms}. Moreover, through $\pi/2$ rotations around $xy$ plane axes, we engineer any $\rr{U}(\phi,\alpha)$. For instance, arbitrary rotations around the $x$ axis are composed as
\begin{equation}
    \rr{U}\!\left(0,\alpha\right)\;\stackrel{z}{=}\;\rr{U}\!\left(\!\frac{\pi}{2}-\alpha,\frac{\pi}{2}\!\right)  \rr{U}\!\left(\!-\frac{\pi}{2},\frac{\pi}{2}\!\right);
\end{equation}
again, generalization to arbitrary $xy$ axes follows directly.\\
\indent Two remarks are important. First, the feature of the SC allowing local control, is that only the interface interacts with an odd number of non blockaded atoms, in turn enabling $\cl{Z}_\text{tot}$. Second, this only works if we never have $\cl{A}-$superatoms close to each other, implying that their distance must always be $\Delta k \geq 4$; otherwise, the intermediate step corrupts the SC.\\
\indent \textit{Entangling gates.---} Placing a $\cl{B}-$superatom between two wires enables the application of a $\rr{CZ}=\mathbb{1}-2\ket{gg}\bra{gg}$ gate, entangling the corresponding qubits. The pulse sequence in Fig.~\ref{fig_3}d is applied when the interfaces are adjacent to the $\cl{B}-$superatom; this induces an effective $2\pi$ rotation of the superatom, conditional on both interfaces being in $\ket{g}$, resulting in a $\rr{CZ}$ gate. The sequence is designed to also induce a $2\pi$ rotation on the $\cl{B}$ atoms: the only byproduct is a global $\cl{Z}_\text{tot}$. The overall operation induced by the pulse in Fig.~\ref{fig_3}d is thus $\rr{CZ}^*=\cl{Z}_\text{tot}\rr{CZ}$. This can be generalized to arbitrary conditional-phase gates, and even multi-qubit gates. 

\indent \textit{Initialization.---} All atoms start in $\ket{g}$. As in Fig.~\ref{fig_1}e, the wires begin with a superatom (in $k=0$); since only these superatoms interact with just one $\cl{B}-$atom, we can apply the protocol for single-qubit gates to bring them all to $\ket{\bar{r}}$. Then, starting the cycle~\eqref{clock} with $\rr{U}_\cl{B}$ results in the SC at $k=2$, with all qubits in $\ket{\psi}=\ket{r}$ after 6 alternating pulses. Subsequently, the transport works as explained, thus the computation starts. Note that we unlock local control at $k=4$.

\textit{Universal arrangement.---} So far, we mapped any algorithm into $\cl{O}(np)$ global pulses driving $\cl{O}(n^2p)$ atoms~\footnote{The depth $p$ is defined in reference to a CM with nearest-neighbor connectivity.}, whose circuit-dependent arrangement mimics the gates~\footnote{These are conservative bounds; for specific circuits consisting of identical parallel gates, as e.g. in variational approaches~\cite{farhi2014quantum, cerezo2021variational}, the scaling is more favorable}. Differently, we now discuss a \emph{universal processor}, consisting of a circuit-independent arrangement of $\cl{O}(n^2)$ atoms; any circuit can then be executed through $\cl{O}(np)$ global pulses. We exploit the fact that by inverting the order in~\eqref{clock}, the interface moves to the left. Then, with arrangements as in Fig.~\ref{fig_3}e, we execute any circuit step-by-step, by moving the information to the left or right to localize it at positions where gates can be triggered. The construction of the universal arrangement works by placing an $\cl{A}-$superatom in each wire, and a $\cl{B}-$superatom between each pair of neighboring wires; these are arranged such that we can execute any gate individually, by moving the interface to the dedicated position (the interface can also be moved  across superatoms  without executing any gate). The number of atoms here no longer depends on $p$, and the arrangement is circuit-independent, realizing a universal processor~\footnote{The total number of atoms needed for processing $n$ qubits is $2n^2 + 3(S+1)n-S$ (see Fig.~\ref{fig_3}e)}.\\
\indent \textit{Suppressing errors.---} While our model incurs a $\cl{O}(n^2)$ overhead, this does not necessarily translate to higher error rates, as quantum information is only stored at the interface, allowing classical correction schemes on the rest of the array. For instance, at stroboscopic times all atoms of one species are in $\ket{g}$, with the interface on the other species; hence inserting species-selective resets allows to periodically eliminate errors on idle atoms. Thus only errors at the interface remain, resulting in a $\cl{O}(n)$ error scaling, similar to the CM. \\
\indent In addition, several techniques can be employed to reduce errors: all steps presented here are designed for didactical simplicity, but could be optimized using optimal control~\cite{PUPILLO_OPTIMAL}. Moreover, specific F\"orster regimes can be chosen to suppress unwanted interactions from Van-der-Waals tails~\cite{supplemental}. Finally, one might incorporate  quantum error correction without local control using measurement-free paradigms~\cite{PhysRevLett.117.130503, perlin2023faulttolerant}, e.g. through 3D atom arrangements ~\cite{barredo2018synthetic}. A detailed analysis of these concepts is left for future work.\\
\indent \textit{Conclusion.---} We showed that the collective many-body dynamics of a static arrangement of globally driven Rydberg atoms is complex enough to enable universal quantum computation. These ideas can be implemented on dual-species Rydberg arrays, but also be transposed to other setups, including single-species arrays (e.g. leveraging hyperfine levels or, potentially, quantum many-body scars~\cite{serbyn2021quantum}), but also other physical platforms. Our model could also be used for hybrid designs, with local control combined with globally controlled modules. Our results open up new avenues for quantum processor design, but also
highlight a new connection between out-of-equilibrium many-body dynamics and quantum information. 

\indent\textit{Acknowledgements.---} We are grateful for valuable discussions with H. Bernien. This research was supported
by the ERC Starting grant QARA (Grant No.~101041435), the Army Research Office (Grant No.~W911NF21-1-0367),  and the European Union’s Horizon 2020 research
and innovation program PASQuanS2 (Grant No.~101079862). FC also acknowledges financial support from Fondazione A. Della Riccia during his visit in Innsbruck.

\bibliography{bibliography}
\end{document}


\preprint{APS/123-QED}
\title{Supplemental material for\\
Universal Quantum Computation in Globally Driven Rydberg Atom Arrays
}

\maketitle

\begin{figure*}[t!]
    \begin{minipage}{0.3\textwidth}
        \includegraphics[width=\textwidth]{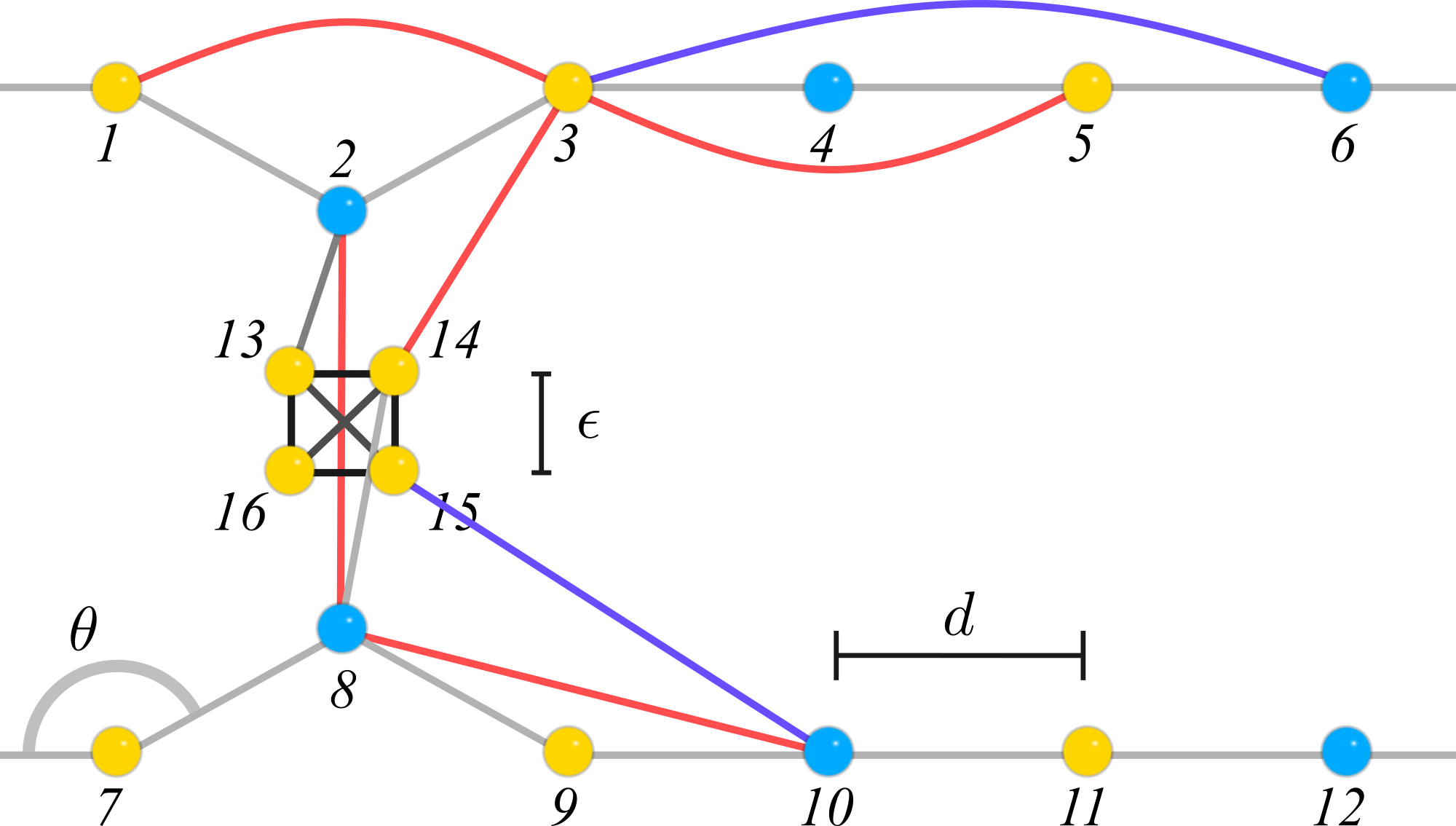}
    \end{minipage}
    \begin{minipage}{0.34\textwidth}
        \includegraphics[width=\textwidth]{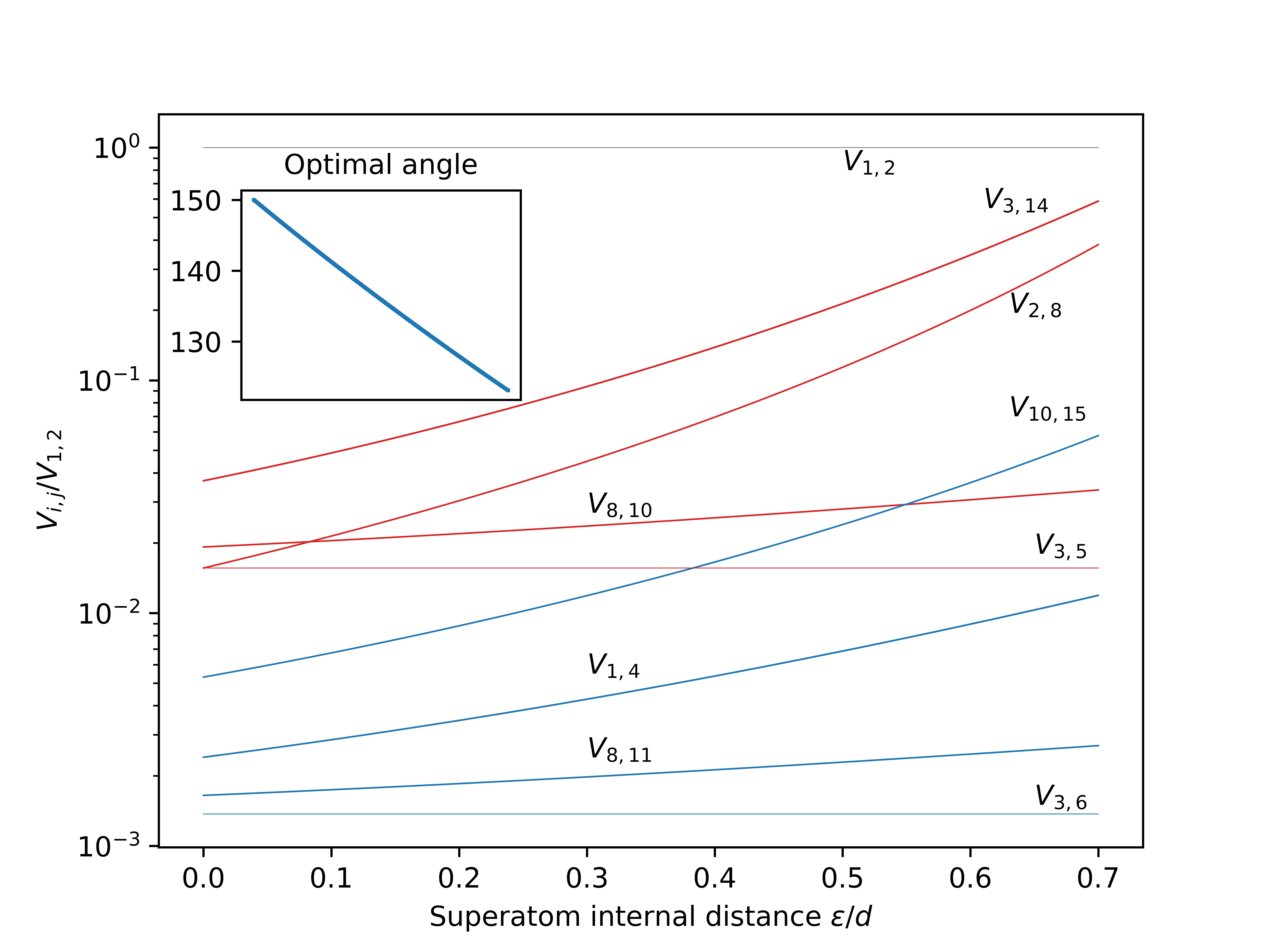}
    \end{minipage}
    \begin{minipage}{0.34\textwidth}
        \includegraphics[width=\textwidth]{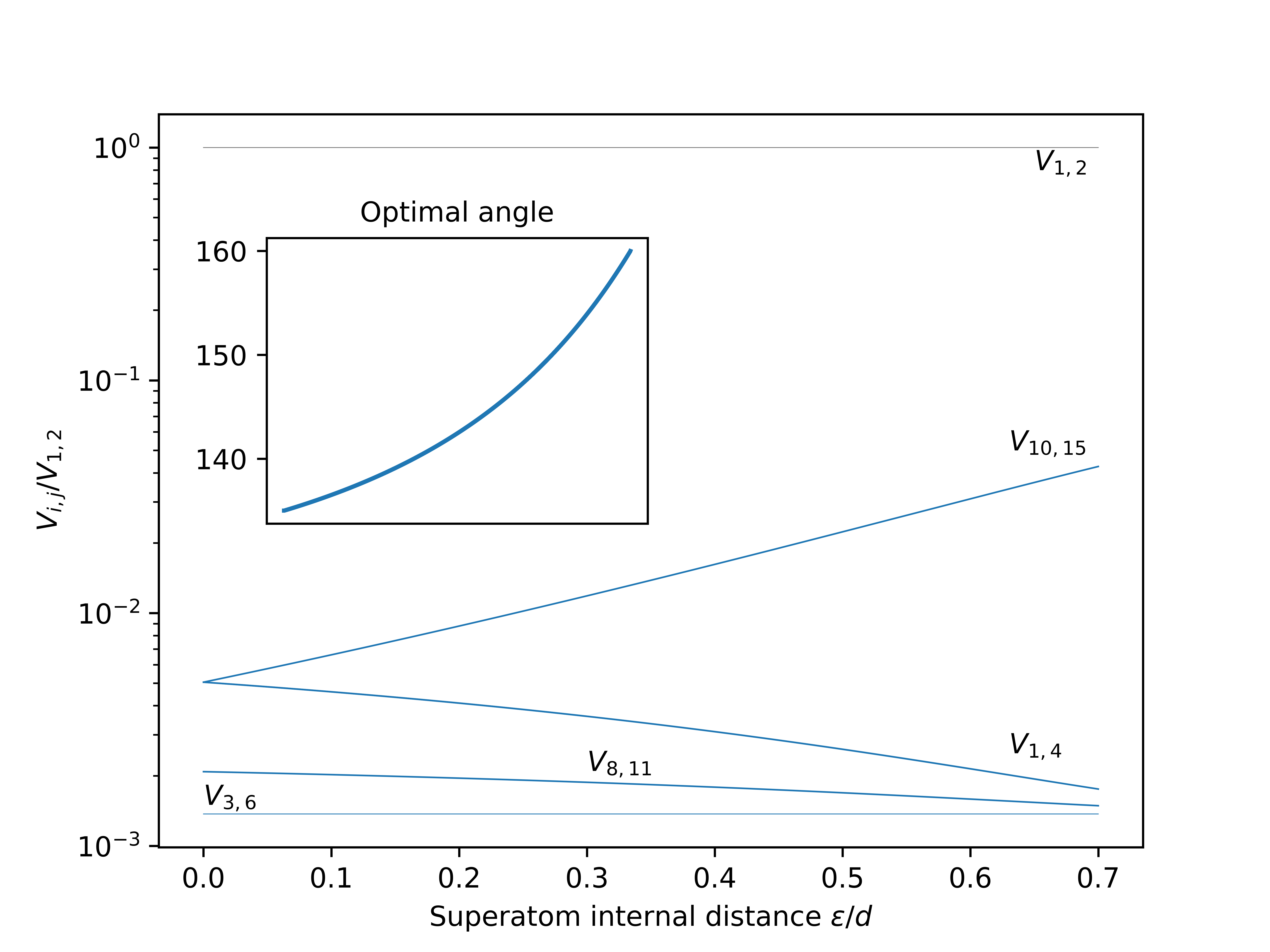}
    \end{minipage}
\caption{To the left, the construction of the $\mathcal{B}-$superatom, which mediates entangling gates between the logical qubits (wires). The wanted, blockade interactions are represented by connecting the atoms with gray lines, while unwanted interactions are depicted in red (intra-species) and blue (inter-species). In the center, the unwanted interactions for different values of $\epsilon$, optimized over all angles $\theta$; here the colors refer to the left figure. To the right, the interacting regime where inter-species interactions are dominating, and intra-species interactions can be neglected.  }
\label{fig_supplemental}
\end{figure*}

\noindent The proposed model for universal quantum computation features a static atomic arrangement, where the dynamics is constrained by the built-in mutual Rydberg blockade interactions. In the main text, we work in the assumption of an ideal blockade regime, corresponding to a dual-species PXP model; this approximation holds when the mutual interaction $V_{i,j}\sim d^{-6}_{i,j}$ between any pair of atoms $i$ and $j$ is always either negligible, or very strong (compared to the Rabi frequency). Here, we justify this approximation by evaluating the strength of the interactions in our setup. Specifically, we consider the construction of $\mathcal{B}-$superatoms for $S=4$, as depicted in Fig.~1 of the main paper; this is potentially the most problematic part of the proposed arrangement. The construction we consider is shown more in detail in Fig.~\ref{fig_supplemental}. Therein, atoms composing the superatom are placed at the vertexes of a square of edge length $\epsilon$, and the arrangement is constructed such that all the wanted interactions occur at most at a distance $d$; specifically, we set
\begin{equation}\label{blockade_distances}
    d_{1,2}=d_{2,3}=d_{3,4}=...=d_{5,6}=d_{6,7}=...=d_{11,12}=d_{2,16}=...=d_{8,15} = d \geq d_{2,13}=...=d_{8,15},
\end{equation}
while unwanted interactions occur at distances $\geq d$. We thus also restrict to $\epsilon \leq d/\sqrt{2}$. Note that given $\epsilon$ all the distances are determined by the angle $\theta$. \\
\indent For quantifying the strength of the blockade, we define 
\begin{equation}
    V^\text{blockade} = \frac{C}{d^6} = V_{1,2}=...
\end{equation}
as the weakest among the wanted interactions (i.e., between the pairs listed in Eq.~\eqref{blockade_distances}); note that e.g. $V_{13,14}\geq V_{3,14}\geq V^\text{blockade}$ in our construction). Moreover, we quantify the violation of the ideal regime through the strongest among all the unwanted interactions,
\begin{align}
    V^\text{unwanted} = \text{max} \left\{ V_{1,3}, \;V_{2,8},\; V_{3,14},\; V_{8,10},\; V_{1,4},\; V_{10,15},\; V_{8,11}  \right\},
\end{align}
where we listed the interactions up to symmetries. In Fig.~\ref{fig_supplemental}, we indicate these unwanted interactions by connecting the atoms with red and blue lines for intra-species and inter-species interactions respectively. Note that $V^\text{unwanted}$ depends on both $\epsilon$ and $\theta$. As a figure of merit for the validity of the PXP approximation, we consider the ratio $V^\text{unwanted}/V^\text{blockade}$; specifically, for given $\epsilon$ and $\theta$ the ideal blockade regime is valid  when the condition
\begin{equation}
    V^\text{unwanted}(\epsilon, \theta)/V^\text{blockade} \ll 1
\end{equation}
holds. Intuitively, small values of $\epsilon$ are advantageous (with ideal superatoms for $\epsilon\longrightarrow 0$), but this parameter is limited by experimental capabilities (and ultimately by fundamental limits); differently, the angle $\theta$ can be chosen arbitrarily. Thus, in what follows for fixed $\epsilon$ we optimize over $\theta$, hence finding the ideal arrangement for given experimental capabilities.\\
\indent First, in the central plot of Fig.~\ref{fig_supplemental} we consider inter- and intra-species interactions at the same level. 
For realistic values of $\epsilon$ such as $\epsilon=0.1$, we find $V^\text{unwanted}/V^\text{blockade}=4.9\times 10^{-2}$, which shows that the ideal blockade appoximation is well justified. It is useful to compare these unwanted interactions arising due to the presence of the superatom, with those that are already present in the 1D wire, the strongest of which being $V_{3,4}\sim 2^{-6}$; in our construction, we have $V^\text{unwanted}/V_{3,4}=2.4$. But even for relatively large values such as $\epsilon=0.4$ one gets $V^\text{unwanted}/V^\text{blockade}=0.14$, which would arguably still justify a PXP approximation.\\
\indent Importantly, F\"orster resonances allow even more convenient interacting regimes; specifically, in a dual-species platform we consider it should be possible to use F\"orster resonances such that the inter-species interactions are strongly dominant over intra-species interactions. 
In such a regime, for instance the dominating unwanted interaction inside a 1D wire would not be the next-nearest interaction $V_{3,4}$, but the third-nearest interaction $V_{4,5}$; in our model, this is extremely convenient because all interactions are actually inter-species, except for those inside the superatoms, which are however not limited by geometric issues of the arrangement, but rather by the non ideal $\epsilon\longrightarrow 0$ limit. In the right-side plot of Fig.~\ref{fig_supplemental}, we plot the strength of unwanted interactions for this regime, where we can effectively neglect the intra-species contributions. Here, even for $\epsilon$ as large as $\epsilon=0.4$ one gets $V^\text{unwanted}/V^\text{blockade}=1.6\times 10^{-2}$, thus largely justifying an ideal blockade approximation; for $\epsilon=0.1$ one even reaches $V^\text{unwanted}/V^\text{blockade}=6.6\times 10^{-3}$. \\
\indent In conclusion, we see that a PXP regime can be fulfilled to an excellent degree with the construction presented in the main text. We also remark that the choice of the F\"orster resonances employed plays a relevant role: even though both the regimes considered allow for an ideal blockade approximation of the many-body dynamics, the second one meets the criteria of the PXP approximation to much a higher degree. Finally, we note that, while in the main text we focus on the case of $S=4$ superatoms, any $S\geq 2$ is suitable for our model; in particular, the case $S=2$ offers very compact and advantageous constructions, which ultimately also lead to a cleaner realization of the PXP approximation.